\documentclass[conference]{IEEEtran}
\usepackage{cite}
\ifCLASSINFOpdf
\else
\fi
\usepackage[cmex10]{amsmath}
\usepackage{array}
\usepackage{mdwmath}
\usepackage{mdwtab}
\usepackage{eqparbox}
\usepackage{amssymb}
\usepackage{graphicx}
\usepackage{mathrsfs}

\newcommand{\Fig}[1]{Fig.~\ref{#1}}
\graphicspath{{figures/eps/}}

\newcommand{\ensemble}{\mathscr{E}}
\newcommand{\field}{{\mathbb{F}}_{q}}

\newtheorem{definition}{Definition}
\newtheorem{lemma}{Lemma}
\newtheorem{theorem}{Theorem}

\newtheorem{remark}{Remark}

\begin{document}

\title{Upper and Lower Bounds on the Minimum Distance of Expander Codes}

\author{\IEEEauthorblockN{Alexey Frolov and Victor Zyablov}
\IEEEauthorblockA{Inst. for Information Transmission Problems\\
Russian Academy of Sciences\\Moscow, Russia\\
Email: \{alexey.frolov, zyablov\}@iitp.ru}}

\maketitle

\begin{abstract}
The minimum distance of expander codes over $GF(q)$ is studied. A new upper bound on the minimum distance of expander codes is derived. The bound is shown to lie under the Varshamov-Gilbert (VG) bound while $q \ge 32$. Lower bounds on the minimum distance of some families of expander codes are obtained. A lower bound on the minimum distance of low-density parity-check (LDPC) codes with a Reed--Solomon constituent code over $GF(q)$ is obtained. The bound is shown to be very close to the VG bound and to lie above the upper bound for expander codes.
\end{abstract}

\IEEEpeerreviewmaketitle

\section{Introduction}
In this work, we consider a family of codes based on expander graphs. The idea of codes on graphs was proposed by Tanner in \cite{Tanner}. Later expander graphs were used by Sipser and Spielman in \cite{SipserSpielman} to obtain asymptotically good codes that can be decoded in time complexity which is linear in the code length. By ``asymptotically good codes'' we mean codes whose rate and relative minimum distance are both bounded away from zero. In \cite{SipserSpielman} both random and explicit constructions of expander graphs were used. The explicit constructions of expander graphs are called Ramanujan graphs and presented in \cite{LubotskyPhilipsSarnak}, \cite{Margulis}. In \cite{Spielman} the construction of expander codes where both encoding and decoding time complexities were linear in the code length was presented. Though the decoder of the Sipser-Spielman construction was guaranteed to correct a number of errors that was a positive fraction of the code length, that fraction was small. Improvements were given in \cite{Zemor}, where the underlying graph was bipartite, and in \cite{SkachekRoth}. In this work, the distance properties of expander codes are studied.
	
There are a lot of works where lower bounds on the minimum distance of families of expander codes are presented (e.g \cite{BargZemor}). In these works the method proposed by Gallager in \cite[Ch.\ 2, pp.\ 13--20]{Gallager} is applied for expander codes. Unfortunately we were unable to find a work where an upper bound was derived. In this work a new upper bound on the minimum distance of expander codes over $GF(q)$ is derived. The bound is shown to lie under the Varshamov-Gilbert bound while $q \ge 32$.

It would seem the results of this work contradict the results of \cite{SkachekRoth_exp}, where a family of codes is presented (they are also called expander codes) which lie close to the Singleton bound if $q$ is large. Nevertheless the constructions are different and the seeming contradiction is the result of some terminology confusion.

\section{Code structure}
Let $G=\left( {{V}_{1}}:{{V}_{2}},E \right)$ be a bipartite undirected connected graph with a vertex set $V={{V}_{1}}\bigcup {{V}_{2}}$ (${{V}_{1}} \bigcap {{V}_{2}}=\emptyset$) and an edge set $E$. Let $\deg \left( {{u}} \right) = {\Delta _1}\,\forall {u} \in {V_1}$, $\deg \left( {{v}} \right) = {\Delta _2}\,\forall {v} \in {V_2}$, $\left| E \right|=n$, then $\left| {{V}_{1}} \right|={{b}_{1}}$, $\left| {{V}_{2}} \right|={{b}_{2}}$, where ${{b}_{1}}=\frac{n}{{{\Delta }_{1}}}$, ${{b}_{2}}=\frac{n}{{{\Delta }_{2}}}$.

Let $\field$ be a Galois field of the power $q$. Let us associate each vertex ${{u}_{i}}\in {{V}_{1}},\,\,i=1,\ldots, {{b}_{1}}$ with a linear $\left( {{\Delta _1},{R_1}{\Delta _1}} \right)$ code $C_{i}^{\left( 1 \right)}$ over $\field$; each vertex ${{v}_{j}}\in {{V}_{2}},\,\,j=1,\ldots, {{b}_{2}}$ with a linear $\left( {{\Delta _2},{R_2}{\Delta _2}} \right)$ code $C_{j}^{\left( 2 \right)}$. Hereinafter $C_{j}^{\left( i \right)}$ will be referred as constituent codes.

For every vertex $u \in V$, we denote by $E\left( u \right)$ the set of edges that are incident with $u$. We assume an ordering on $E$. For a word $\mathbf{z}={{\left( {{z}_{e}} \right)}_{e\in E}}$(whose entries are indexed by $E$), we denote by ${{\left( \mathbf{z} \right)}_{E\left( u \right)}}$ the sub-block of $\mathbf{z}$ that is indexed by $E\left( u \right)$.

Now we are ready to give a definition of an expander code:

\begin{definition}
$C$ is an expander code if
\setlength{\arraycolsep}{0.0em}
\begin{eqnarray*}
 C = \left\{{\bf{c}} \in \field^{\left| E \right|} : \left( {{\left( {\bf{c}} \right)}_{E\left( {{u_i}} \right)}} 	\in C_i^{\left( 1 \right)}\,\,\forall {u_i} \in {V_1} \right) \wedge \right. \\
 \left. \left( {{\left( {\bf{c}} \right)}_{E\left( {{v_j}} \right)}} \in C_j^{\left( 2 \right)}\,\,\forall {v_j} \in {V_2} \right) \right\}
\end{eqnarray*}
\setlength{\arraycolsep}{5pt}
\end{definition}

$C$ is a linear code so there is a parity-check matrix corresponding to it. Let $\mathbf{H}_{i}^{\left( 1 \right)}$ be a parity-check matrix of a constituent code $C_{i}^{\left( 1 \right)}$,  $\mathbf{H}_{j}^{\left( 2 \right)}$ be a parity-check matrix of a constituent code $C_{j}^{\left( 2 \right)}$, then a parity-check matrix $\mathbf{H}$ corresponding to code $C$ is:

\begin{equation}
{\bf{H}} = \left( {\begin{array}{c}
   {{\pi _1}\left( {diag\left( {{\bf{H}}_1^{\left( 1 \right)}, {\bf{H}}_2^{\left( 1 \right)}, \ldots, {\bf{H}}_{{b_1}}^{\left( 1 \right)}} \right)} \right)}  \\
   {{\pi _2}\left( {diag\left( {{\bf{H}}_1^{\left( 2 \right)}, {\bf{H}}_2^{\left( 2 \right)}, \ldots, {\bf{H}}_{{b_2}}^{\left( 2 \right)}} \right)} \right)}  \\
\end{array}} \right),
\label{H_exp}
\end{equation}

where

\begin{eqnarray*}
&{}{}&diag\left( {{\bf{H}}_1^{\left( i \right)}, {\bf{H}}_2^{\left( i \right)}, \ldots, {\bf{H}}_{{b_i}}^{\left( i \right)}} \right) \\
 &&  {=} {\left( {\begin{array}{cccc}
   {\bf{H}}_1^{\left( i \right)} & {\bf{0}} &  \cdots  & {\bf{0}}  \\
   {\bf{0}} & {\bf{H}}_2^{\left( i \right)} &  \cdots  & {\bf{0}}  \\
    \vdots  &  \ddots  &  \ddots  &  \vdots   \\
   {\bf{0}} & {\bf{0}} &  \cdots  & {\bf{H}}_{{b_i}}^{\left( i \right)}  \\
\end{array}} \right)_{\left( {1 - {R_i}} \right)n \times n}}, \\
\end{eqnarray*}

${{\pi }_{i}}$ is a column permutation of $diag\left( {\bf H}_{1}^{\left( i \right)}, {\bf H}_{2}^{\left( i \right)}, \ldots, {\bf H}_{{{b}_{i}}}^{\left( i \right)} \right)$ uniquely defined by a graph $G$ and by a fixed order on $E$.

\begin{remark}
The size of ${\bf{H}}$ is ${\left( {\left( {1 - {R_1}} \right) + \left( {1 - {R_2}} \right)} \right)n \times n}$.
\end{remark}

Now we will determine the parameters of the obtained code. The length of $C$ is equal to $\left| E \right|=n$, the rate of $C$ is
\setlength{\arraycolsep}{0.0em}
\begin{equation}
R \geq {{R}_{1}} + {{R}_{2}} - 1
\label{rate}
\end{equation}
\setlength{\arraycolsep}{5pt}
The equality takes place in case of full rank of ${\bf H}$.

\section{New upper bound}

We will use the method similar to the method from \cite{litsyn}. Let $C'$ be an expander code. Its parity-check matrix is given by (\ref{H_exp}). Without loss of generality ${{R}_{1}} \le {{R}_{2}}$. The parity-check matrix of $C'$ can be transformed to the form:
\begin{equation*}
{\bf{H}} = \left( {\begin{array}{c}
   diag\left( {\bf{H}}_1^{\left( 1 \right)}, {\bf{H}}_2^{\left( 1 \right)}, \ldots, {\bf{H}}_{b_1}^{\left( 1 \right)} \right)  \\
   \pi _1^{ - 1}{\pi _2}\left( {diag\left( {\bf{H}}_1^{\left( 2 \right)}, {\bf{H}}_2^{\left( 2 \right)}, \ldots , {\bf{H}}_{b_2}^{\left( 2 \right)} \right)} \right)  \\
\end{array}} \right).
\end{equation*}

Let $C$ be a code corresponding to ${\bf H}$. Codes $C$ and $C'$ are equivalent hence they have the same distance properties. Now we are ready to prove the theorem:

\begin{theorem}
Let $C$ be an expander code, then
\begin{equation*}
d\left( C \right)\le \underset{{{b}_{1}}\ge b'\ge \frac{\left( R_1 - R \right)}{{{R}_{1}}}b_1+\frac{1}{{{R}_{1}}{{\Delta }_{1}}}}{\mathop{\min }}\,\left\{ \frac{{{q}^{\widetilde{k}-1}}\left( q-1 \right)}{{{q}^{\widetilde{k}}}-1}b'{{\Delta }_{1}} \right\},
\end{equation*}
where $\widetilde{k}=b'{{R}_{1}}{{\Delta }_{1}}-\left( R_1 - R \right)n$, $b'\in \mathbb{N}$.
\end{theorem}

\begin{IEEEproof}
Let us consider a code $\widetilde{C}$ of length $\widetilde{n}=b'{{\Delta }_{1}}$, $b'\in \mathbb{N}$. The parity-check matrix $\widetilde{\bf H}$ of the code is shown in \Fig{subcode}. The code $\widetilde{C}$ correspond to a subcode $C''$ of $C$. We just need to add a prefix of $n - \widetilde{n}$ zeros to the word $\widetilde{\bf c}$ of $\widetilde{C}$ to obtain the word ${\bf c''}$ of $C''$, i.e.

\begin{equation*}
{\bf c''} = \left( \bf{0} \: \widetilde{\bf c} \right).
\end{equation*}

Hence,

\begin{equation*}
d\left( C \right) \leq d \left( C'' \right) = d \left( \widetilde{C} \right)
\end{equation*}

\begin{figure}[!t]
\centering
\includegraphics[width=\linewidth]{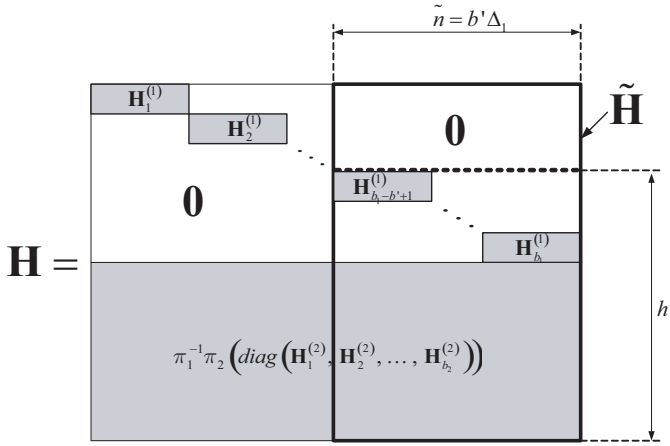}
\caption{Parity-check matrix of $\widetilde{C}$}
\label{subcode}
\end{figure}

Let us consider the code $\widetilde{C}$ in more detail. The height $h$ of its parity-check matrix and hence the number of check symbols in $\widetilde{C}$ are upper bounded with $\left( {{R_1} - {R}} \right)n + b'\left( {1 - {R_1}} \right){\Delta _1}$, so the dimension $\widetilde{k}$ of $\widetilde{C}$ can be estimated as follows

\begin{equation*}
\widetilde{k} \geq b'{R_1}{\Delta _1} - \left( {R_1} - {R} \right)n
\end{equation*}

For the condition $\widetilde{k} \geq 1$ to be satisfied the following condition is sufficient

\begin{equation*}
b'{R_1}{\Delta _1} - \left( {R_1} - {R} \right)n \geq 1
\end{equation*}

So we have such a condition

\begin{equation*}
b' \geq \frac{R_1 - R}{R_1}b_1 + \frac{1}{R_1 \Delta_1}
\end{equation*}

After applying the Plotkin bound to $\widetilde{C}$ we obtain the needed result

\begin{equation*}
d\left( C \right)\le \underset{b'}{\mathop{\min }}\,\left\{ \frac{{{q}^{\widetilde{k}-1}}\left( q-1 \right)}{{{q}^{\widetilde{k}}}-1}b'{{\Delta }_{1}} \right\},
\end{equation*}
where $b'$ satisfy the condition $\frac{\left( R_1 - R \right)}{{{R}_{1}}}b_1+\frac{1}{{{R}_{1}}{{\Delta }_{1}}} \le b'\le {{b}_{1}}$.

\end{IEEEproof}

\begin{remark}
In fact we can apply the stronger bound (e.g. the Elias--Bassalygo bound or the MRRW bound) and obtain a tighter bound, but even the Plotkin bound is enough for our purpose.
\end{remark}

Now we will derive an asymptotic form of the new bound.

\begin{theorem}
Let $\left\{ {{C}_{i}} \right\}_{i=1}^{\infty }$ be a sequence of expander codes with rates $R\left( {{C}_{i}} \right)=R$ and lengths $n\left( {{C}_{i}} \right)=i\times\mathop{LCM}\left( {{\Delta }_{1}},{{\Delta }_{2}} \right)$\footnote{by $LCM(a, b)$ we mean least common multiplier of $a$ and $b$, i.e. $LCM(a, b) = \mathop{min}\limits_m \left\{ m : (a | m) \wedge (b | m) \right\}$} then

\begin{equation*}
\delta = \lim_{i\to \infty }{\left( \frac{d\left( {{C}_{i}} \right)}{n\left( {{C}_{i}} \right)} \right)} \le \frac{q-1}{q}\left( \frac{1-R}{1+R} \right)
\end{equation*}
\end{theorem}

\begin{IEEEproof}
Let us choose
\begin{equation*}
b' = \left\lceil {b_1\left( {\frac{{R_1 - R}}{{R_1}}} \right)} \right\rceil + f\left( n \right),
\end{equation*}
where $f\left( n \right)\to \infty $ while $n\to \infty $ and $f\left( n \right)=o\left( n \right)$, then

\setlength{\arraycolsep}{0.0em}
\begin{eqnarray*}
{d\left( C \right)}&{}\le{}&\frac{{{q^{{R_1}{\Delta _1}f\left( n \right) - 1}}\left( {q - 1} \right)}}{{{q^{{R_1}{\Delta _1}f\left( n \right)}} - 1}}\\
&{\times}&\:\left( {\left( {\frac{{R_1 - R}}{{R_1}}} \right)n + \left( {f\left( n \right) + 1} \right){\Delta _1}} \right)
\end{eqnarray*}
\setlength{\arraycolsep}{5pt}

After dividing on $n$ and taking the limit we have
\begin{equation}
\delta \le \frac{q-1}{q}\left( \frac{R_1 - R}{R_1} \right).
\label{delta_bound}
\end{equation}

Finally, from conditions $R_1 \leq R_2$ and (\ref{rate}) we have
\begin{equation*}
R_1 \leq \frac{R_1 + R_2}{2} \leq \frac{1+R}{2}
\end{equation*}
and after substituting it to (\ref{delta_bound}) we obtain the needed result
\begin{equation*}
\delta \le \frac{q-1}{q}\left( \frac{1 - R}{1 + R} \right).
\end{equation*}

\end{IEEEproof}

\section{Lower bounds}

In this section, we obtain the lower bounds on the minimum distance for three code ensembles. Let us introduce needed notations and prove statements common for all the ensembles.

Let $\ensemble$ be an ensemble of codes of length $n$. By ${\overline {A\left( W \right)} }$ we denote a number of code words of weight $W$ in a code averaged over the ensemble, i.e.
\begin{equation*}
\overline{A\left( W \right)}=\frac{1}{\left| \ensemble \right|}\sum\limits_{i=1}^{\left| \ensemble \right|}{{{A}_{i}}\left( W \right)},
\end{equation*}
where ${{A_i}\left( W \right)}$ is a number of code words of weight $W$ in a code $C_i \in \ensemble$.

\begin{theorem}
If the condition
\begin{equation}
\sum\limits_{W=1}^{d}{\overline{A\left( W \right)}}<1
\label{sum_A_W}
\end{equation}
is satisfied for $\ensemble$ then there exist a code $C \in \ensemble : d\left( C \right)>d$.
\end{theorem}
\begin{IEEEproof}
$\sum\limits_{W = 1}^d {\overline {A\left( W \right)} }  < 1 \Rightarrow \sum\limits_{W = 1}^d {\sum\limits_{i = 1}^{\left| {{\ensemble}} \right|} {{A_i}\left( W \right)} }  < \left| {{\ensemble}} \right|$, which means that the total number of code words of small weight in $\ensemble$ is less than the number of codes in $\ensemble$, therefore there exist a code $C \in \ensemble$ which does not contain the words:
\begin{equation*}
d\left( C \right)>d
\end{equation*}
\end{IEEEproof}

\begin{remark}
Note, that
\begin{equation}
\overline {A\left( W \right)}  = \frac{1}{{\left| \ensemble \right|}}\sum\limits_{i = 1}^{\left| \ensemble \right|} {{A_i}\left( W \right)}  = \frac{1}{{\left| \ensemble \right|}}\sum\limits_{j = 1}^{\left| {{V_W}} \right|} {N\left( \ensemble, {{\bf{v}}_j^{\left( W \right)}} \right)},
\label{A_W_calc}
\end{equation}
where ${{V}_{W}} = \left\{ {\bf v^{\left( W \right)}} \in \field^n : \left\| {\bf{v}}^{\left( W \right)} \right\| = W \right\}$ ($\left\| \bf v \right\|$ is the Hamming weight of $\bf v$), $N\left( \ensemble, {{{\bf{v}}}} \right)$ is a number of codes from $\ensemble$ containing ${\bf{v}}$ as a code word.

\end{remark}
	
Now consider some particular code ensembles.

\subsection{Ensemble $\ensemble_1\left( {{\Delta _0},b} \right)$ of expander codes with a Reed--Solomon constituent code}
Let us consider a block-diagonal matrix
\begin{equation}
{{\bf{H}}_b} = {\left( {\begin{array}{cccc}
   {{{\bf{H}}_0}} & {\bf{0}} &  \cdots  & {\bf{0}}  \\
   {\bf{0}} & {{{\bf{H}}_0}} &  \cdots  & {\bf{0}}  \\
    \vdots  &  \ddots  &  \ddots  &  \vdots   \\
   {\bf{0}} & {\bf{0}} &  \cdots  & {{{\bf{H}}_0}}  \\
\end{array}} \right)_{\left( {1 - {R_0}} \right)n \times n}},
\label{H_b}
\end{equation}
where ${\bf{H}}_0$ is a parity-check matrix of a $\left( {{\Delta _0},{R_0}{\Delta _0}} \right)$ Reed--Solomon code over $\field$, $n = \Delta_0 b$. By ${\varphi}\left( {{{\bf{H}}_b}} \right)$ we denote the matrix obtained from ${\bf{H}}_b$ by an arbitrary permutation of columns and multiplying them by arbitrary nonzero elements of $\field$. Then the matrix
\begin{equation*}
{\bf{H}} = {\left( {\begin{array}{c}
   {{\varphi _1}\left( {{{\bf{H}}_b}} \right)}  \\
   {{\varphi _2}\left( {{{\bf{H}}_b}} \right)}  \\
\end{array}} \right)_{2\left( {1 - {R_0}} \right)n \times n}}
\end{equation*}
constructed using two matrices as layers, is a sparse parity-check matrix of a code from $\ensemble_1\left( {{\Delta _0},b} \right)$.

We define an ensemble $\ensemble_1\left( {{\Delta _0},b} \right)$ as follows:
\begin{definition}
Elements of the ensemble $\ensemble_1\left( {{\Delta _0},b} \right)$ are obtained by independent choice of permutations ${{\pi }_{i}}$ and nonzero constants ${{c}_{i,j}}, i=1,2; \: j=1,2, \ldots, n$, by which parity-check matrices of layers are multiplied.
\end{definition}

\begin{remark}
Each code from ${\ensemble_1}\left( {{\Delta _0},b} \right)$ is an expander code ($\Delta_1=\Delta_2=\Delta_0$, ${{R}_{1}}={{R}_{2}}={{R}_{0}}$), therefore the upper bound is valid for all of them.
\end{remark}

\begin{remark}
$\left| \ensemble_1\left( {{\Delta _0},b} \right) \right| = {\left( {n!{(q-1)^n}} \right)^2}$
\end{remark}

\begin{lemma}
A number of code words of weight $W$ in a code averaged over the ensemble $\ensemble_1\left( {{\Delta _0},b} \right)$
\begin{equation*}
\overline {A\left( W \right)}  = \frac{{{{\left( {{A_1}\left( W \right)} \right)}^2}}}{{{{\left( {q - 1} \right)}^W}\left( {\begin{array}{c}
   n  \\
   W  \\
\end{array}} \right)}},
\end{equation*}
where ${{A}_{1}}\left( W \right)$ is a number of code words of weight $W$ in the first layer.
\end{lemma}
\begin{IEEEproof}
Consider a fixed vector ${{\bf{v}}^{\left( W \right)}}$ of length $n$, $\left\| {{{\bf{v}}^{\left( W \right)}}} \right\| = W$. In accordance to equation (\ref{A_W_calc}) we need to calculate $N\left( { \ensemble_1, {{\bf{v}}^{\left( W \right)}}} \right)$. Now we consider the ensembles of first ($L_1$) and second ($L_2$) layers separately. If we know the number of layers from $L_1$ containing ${{\mathbf{v}}^{\left( W \right)}}$ as a code word $\left( {N}\left( {L_1, {{\bf{v}}^{\left( W \right)}}} \right) \right)$ and the number of layers from $L_2$ containing ${{\mathbf{v}}^{\left( W \right)}}$ as a code word $\left( {N}\left( {L_2, {{\bf{v}}^{\left( W \right)}}} \right) \right)$, then
\begin{equation*}
N\left( { \ensemble_1, {{\bf{v}}^{\left( W \right)}}} \right) = {N}\left( {L_1, {{\bf{v}}^{\left( W \right)}}} \right){N}\left( {L_2, {{\bf{v}}^{\left( W \right)}}} \right),
\end{equation*}
it follows from the fact that permutations and nonzero elements are chosen independently. For the same reason $L_1 = L_2$, hence
\begin{equation}
N\left( { \ensemble_1, {{\bf{v}}^{\left( W \right)}}} \right) = {\left( {{N}\left( {L_1, {{\bf{v}}^{\left( W \right)}}} \right)} \right)^2}.
\label{N}
\end{equation}

To calculate ${N}\left( {L_1, {{\bf{v}}^{\left( W \right)}}} \right)$, we proceed as follows: fix a permutation $\pi_1$, and perform permutations and multiplications by constants over elements of a vector ${{\bf{v}}^{\left( W \right)}}$ but not over columns of a parity-check matrix. Clearly, these problems are equivalent, and since nothing depends on a particular permutation $\pi_1$, we let it be the identity permutation.

In accordance to properties of ${{\varphi }_{i}}$ there are all possible vectors of weight $W$ among $\left\{ {{\varphi }_{i}}^{-1}\left( {{\mathbf{v}}^{\left( W \right)}} \right) \right\}_{i=1}^{{\left| L_1 \right|}}$ and each of them is repeated $K$ times, where
\begin{equation*}
K=W!\left( n-W \right)!{{\left( q-1 \right)}^{n-W}}.
\end{equation*}

Thus, we obtain
\setlength{\arraycolsep}{0.0em}
\begin{eqnarray*}
{N}\left( {L_1, {{\bf{v}}^{\left( W \right)}}} \right) &{}={}& {{A}_{1}}\left( W \right)K \\
&{}={}& {{A}_{1}}\left( W \right)W!\left( n-W \right)!{{\left( q-1 \right)}^{n-W}}.
\end{eqnarray*}
\setlength{\arraycolsep}{5pt}

And finally,
\begin{equation*}
N\left( { \ensemble_1, {{\bf{v}}^{\left( W \right)}}} \right) = {{\left( {{A}_{1}}\left( W \right)W!\left( n-W \right)!{{\left( q-1 \right)}^{n-W}} \right)}^{2}}.
\end{equation*}

One can notice that the value $N\left( { \ensemble_1, {{\bf{v}}^{\left( W \right)}}} \right)$ is the same for all the vectors of weight $W$, hence in accordance to (\ref{A_W_calc}) we obtain the needed result.
\end{IEEEproof}

In the next lemma we obtain an upper bound on ${\overline {A\left( W \right)} }$.

\begin{lemma} A number of code words of weight $W$ in a code averaged over the ensemble $\ensemble_1\left( {{\Delta _0},b} \right)$ can be estimated as follows
\begin{equation*}
\overline{A\left( W \right)} \leqslant {{q}^{-nF_1\left( \delta ,{{\Delta }_{0}} \right)}},
\end{equation*}
where
\setlength{\arraycolsep}{0.0em}
\begin{eqnarray*}
 F_1\left( {\delta ,{\Delta _0}} \right) &{}={}& {h_q}\left(
 \delta  \right) + \delta {\log _q}\left( {q - 1} \right) \\
 &{+}&\:  2 \mathop {\max }\limits_{s > 0} \left( {\delta {{\log }_q}\left( s \right) - \frac{1}{{{\Delta _0}}}\log_q\left( {{g_0}\left( {s,{\Delta _0}} \right)} \right)} \right),
\end{eqnarray*}
\setlength{\arraycolsep}{5pt}
\\
$\delta =\frac{W}{n}$, ${{h}_{q}}\left( \delta  \right)=-\delta {{\log }_{q}}\left( \delta  \right)-\left( 1-\delta  \right){{\log }_{q}}\left( 1-\delta  \right)$ -- $q$-ry entropy function and ${{g}_{0}}\left( s,{{\Delta }_{0}} \right)$ is a generating function of weights of code words of constituent code.
\end{lemma}

\begin{IEEEproof}
Note that in each layer the sets of positions occupied by code symbols of constituent codes are disjoint. At the same time, all positions are covered; hence, the generating function of layer $G\left( s \right)$ looks like:
\begin{equation*}
G\left( s \right)={{g}_{0}}^{\frac{n}{{{\Delta }_{0}}}}\left( s,{{\Delta }_{0}} \right),
\end{equation*}
then
\begin{equation*}
{{A}_{1}}\left( W \right)=\left[ {{s}^{W}} \right]\left( {{g}_{0}}^{\frac{n}{{{\Delta }_{0}}}}\left( s,{{\Delta }_{0}} \right) \right)
\end{equation*}
After using an evident estimation
\begin{equation*}
{{A}_{1}}\left( W \right)\le \underset{s>0}{\mathop{\min }}\,\left( \frac{{{g}_{0}}^{\frac{n}{{{\Delta }_{0}}}}\left( s,{{\Delta }_{0}} \right)}{{{s}^{W}}} \right)
\end{equation*}
we obtain
\begin{equation*}
\overline{A\left( W \right)}\le {{q}^{-nF_1\left( \delta ,{{\Delta }_{0}} \right)}},
\end{equation*}
where $\delta =\frac{W}{n}$, ${{h}_{q}}\left( \delta  \right)=-\delta {{\log }_{q}}\left( \delta  \right)-\left( 1-\delta  \right){{\log }_{q}}\left( 1-\delta  \right)$,
\setlength{\arraycolsep}{0.0em}
\begin{eqnarray*}
 F_1\left( {\delta ,{\Delta _0}} \right) &{}={}& {h_q}\left(
 \delta  \right) + \delta {\log _q}\left( {q - 1} \right) \\
 &{+}&\:  2 \mathop {\max }\limits_{s > 0} \left( {\delta {{\log }_q}\left( s \right) - \frac{1}{{{\Delta _0}}}\log_q\left( {{g_0}\left( {s,{\Delta _0}} \right)} \right)} \right).
\end{eqnarray*}
\setlength{\arraycolsep}{5pt}
\end{IEEEproof}
\begin{remark}
The generating function of weights of code words of a $\left( \Delta_0, R_0 \Delta_0\right)$ Reed--Solomon code can be estimated as follows:
\begin{equation*}
{{g}_{0}}\left( s,{{\Delta }_{0}} \right) \le 1 + \sum\limits_{i = {d_0}}^{{\Delta _0}} {\left( {\left( {\begin{array}{c}
   \Delta_0  \\
   i  \\
\end{array}} \right){\left(q-1\right)^{i - {d_0} + 1}}} s^i\right)},
\end{equation*}
where $d_0 = \left( 1-R_0 \right)\Delta_0 + 1$.
\end{remark}

\begin{theorem}
If there exist at least one positive root (with respect to unknown $\delta$) of equation
\begin{equation}
F_1\left( {\delta ,{\Delta _0}} \right) = 0
\label{funct_1}
\end{equation}
then in the ensemble $\ensemble_1\left( {{\Delta _0},b} \right)$ there exist codes $\left\{ {{C_i}} \right\}_{i = 1}^{N(b)}$ $\left( \mathop {\lim }\limits_{b \to \infty }{\frac{N(b)}{\left| \ensemble_1\left( {{\Delta _0},b} \right) \right|}} = 1 \right)$ such that $d \left( C_i \right) \ge \left( \delta_1 - \varepsilon\right) n$, where $\varepsilon$ is an arbitrary small positive number; $\delta_1$ is a positive root of equation (\ref{funct_1}).
\end{theorem}
\begin{IEEEproof}
We just need to prove that
\begin{equation*}
\mathop {\lim }\limits_{n \to \infty } \left( {\sum\limits_{W = 1}^{\left\lfloor {\left( {{\delta _1} - \varepsilon } \right)n} \right\rfloor } {\overline {A\left( W \right)} } } \right) = 0.
\end{equation*}
The proof is similar to the proof of Theorem 2 in \cite{FrolovZyablov}. We omit the proof here.
\end{IEEEproof}

\subsection{Ensemble $\ensemble_2\left( {{\Delta _0},b} \right)$ of expander codes with a constituent code from an expurgated ensemble of random codes}

In previous section we use a Reed--Solomon code as a constituent code. Unfortunately the length of this code can't be sufficiently large ($\Delta_0 \leq q + 1$). In this section we will choose a constituent code from an expurgated ensemble of random codes and use it as a constituent code. In this case we don't have any constraints on the constituent code length.

\begin{theorem}
For each $\Delta_0$ and $R_0$ there exist a linear $\left( \Delta_0, R_0 \Delta_0 \right)$ code $C_0$ with such a spectrum
\begin{enumerate}
\item {$A_0 \left( 0 \right) = 1$;}
\item {$A_0 \left( W \right) \leq 2 \Delta_0 \left( {\begin{array}{*{20}{c}}
   {{\Delta _0}}  \\
   W  \\
\end{array}} \right){{\left( {q - 1} \right)}^W}{q^{ - {\Delta _0}\left( {1 - {R_0}} \right)}}$\\ for $W \in \left[ 1, \Delta_0 \right]$.}
\end{enumerate}
\end{theorem}
\begin{IEEEproof}
The proof can be found in \cite[Ch.\ 2, Th.\ 2.4]{BlockZyablov}.
\end{IEEEproof}

The generating function of weights of code words of $C_0$ can be estimated as follows:
\setlength{\arraycolsep}{0.0em}
\begin{eqnarray*}
{{g}_{0}}\left( s,{{\Delta }_{0}} \right) &{}\le{}& 1 \\
&{}+{}& \sum\limits_{i = 1}^{{\Delta _0}} {\left( \left\lfloor 2\Delta_0 {\left( {\begin{array}{c}
   \Delta_0  \\
   i  \\
\end{array}} \right){\left(q-1\right)^{i}}} q^{- \Delta_0 \left( 1 - R_0 \right)} \right\rfloor s^i\right)},
\end{eqnarray*}
\setlength{\arraycolsep}{5pt}

All the proofs here are analogical to the proofs for $\ensemble_1\left( {{\Delta _0},b} \right)$. We will just give the main result.

\begin{theorem}
If there exist at least one positive root (with respect to unknown $\delta$) of equation
\begin{equation}
F_2\left( {\delta ,{\Delta _0}} \right) = 0
\label{funct_2}
\end{equation}
then in the ensemble $\ensemble_2\left( {{\Delta _0},b} \right)$ there exist codes $\left\{ {{C_i}} \right\}_{i = 1}^{N(b)}$ $\left( \mathop {\lim }\limits_{b \to \infty }{\frac{N(b)}{\left| \ensemble_2\left( {{\Delta _0},b} \right) \right|}} = 1 \right)$ such that $d \left( C_i \right) \ge \left( \delta_2 - \varepsilon\right) n$, where $\varepsilon$ is an arbitrary small positive number; $\delta_2$ is a positive root of equation (\ref{funct_2}),
\setlength{\arraycolsep}{0.0em}
\begin{eqnarray*}
 F_2\left( {\delta ,{\Delta _0}} \right) &{}={}& \left( {{h_q}\left( \delta  \right) + \delta {{\log }_q}\left( {q - 1} \right)} \right) \\
 &+&\: \mathop {2  {\mathop {\max }}}\limits_{s > 0} \left( {\delta {{\log }_q}\left( s \right) - \frac{1}{{{\Delta _0}}}\log_q\left( {{g_0}\left( {s,{\Delta _0}} \right)} \right)} \right).
\end{eqnarray*}
\setlength{\arraycolsep}{5pt}
\end{theorem}

\subsection{Ensemble $\ensemble_3\left( {{\Delta _0},b} \right)$ of non-binary LDPC codes with a Reed--Solomon constituent code}

Let us consider the matrix
\begin{equation*}
{\bf{H}} = {\left( {\begin{array}{c}
   {{\varphi _1}\left( {{{\bf{H}}_b}} \right)}  \\
   {{\varphi _2}\left( {{{\bf{H}}_b}} \right)}  \\
    \vdots   \\
   {{\varphi _\ell }\left( {{{\bf{H}}_b}} \right)}  \\
\end{array}} \right)_{\ell \left( {1 - {R_0}} \right)n \times n}}
\end{equation*}
\\
constructed using $\ell$ layers, the notion ${\varphi}\left( {{{\bf{H}}_b}} \right)$ was introduced while defining of $\ensemble_1\left( {{\Delta _0},b} \right)$ ensemble. The matrix is a sparse parity-check matrix of a code from $\ensemble_3\left( {{\Delta _0},b} \right)$.

\begin{definition}
Elements of the ensemble $\ensemble_3\left( {{\Delta _0},b} \right)$ are obtained by independent choice of permutations ${{\pi }_{i}}$ and nonzero constants ${{c}_{i,j}}, i=1,2, \ldots, \ell; \, j=1,\,2,\ldots ,n$, by which parity-check matrices of layers are multiplied.
\end{definition}

\begin{remark}
The definition is similar to the definition of ensemble $\ensemble_1\left( {{\Delta _0},b} \right)$ but the parity-check matrices here consist of $\ell$ layers rather than $2$ ones.
\end{remark}
\begin{remark}
The codes are not expander codes and hence the upper bound is not valid for them. They are given here for comparison with expander codes.
\end{remark}

All the proofs here are analogical to the proofs for $\ensemble_1\left( {{\Delta _0},b} \right)$. We will just give the main result.

\begin{theorem}
If there exist at least one positive root (with respect to unknown $\delta$) of equation
\begin{equation}
F_3\left( {\delta ,{\Delta _0}} \right) = 0
\label{funct_3}
\end{equation}
then in the ensemble $\ensemble_3\left( {{\Delta _0},b} \right)$ there exist codes $\left\{ {{C_i}} \right\}_{i = 1}^{N(b)}$ $\left( \mathop {\lim }\limits_{b \to \infty }{\frac{N(b)}{\left| \ensemble_3\left( {{\Delta _0},b} \right) \right|}} = 1 \right)$ such that $d \left( C_i \right) \ge \left( \delta_3 - \varepsilon\right) n$, where $\varepsilon$ is an arbitrary small positive number; $\delta_3$ is a positive root of equation (\ref{funct_3}),
\setlength{\arraycolsep}{0.0em}
\begin{eqnarray*}
 F_3\left( {\delta ,{\Delta _0}} \right) &{}={}& \left( {\ell  - 1} \right)\left( {{h_q}\left( \delta  \right) + \delta {{\log }_q}\left( {q - 1} \right)} \right) \\
 &+&\: \mathop {\ell  {\mathop {\max }}}\limits_{s > 0} \left( {\delta {{\log }_q}\left( s \right) - \frac{1}{{{\Delta _0}}}\log_q\left( {{g_0}\left( {s,{\Delta _0}} \right)} \right)} \right).
\end{eqnarray*}
\setlength{\arraycolsep}{5pt}
\end{theorem}

\section{Numerical results}
Results obtained for $q=64$ and $q=1024$ are shown in Tables \ref{t_64} and \ref{t_1024}, respectively. The result for $\ensemble_1$, $\ensemble_2$ and $\ensemble_3$ are maximized over $\Delta_0$. Note that $\Delta_0 \le q + 1$ for $\ensemble_1$ and $\ensemble_3$. The derived upper bound lies below the Varshamov-Gilbert bound when $R \in \left(0.25; 0.89 \right)$ for $q=64$. This interval is widening while $q$ is growing. For $q = 1024$ we have such an interval $\left(0.05; 0.99 \right)$.

\begin{table}[!t]
\renewcommand{\arraystretch}{1.3}
\caption{Results for $q=64$}
\label{t_64}
\centering
\begin{tabular}{|c||c|c|c|c|l|}
\hline
R & VG & Upper & $\delta_1\left(R\right); \Delta_0$ & $\delta_2\left(R\right); \Delta_0$ & $\delta_3\left(R\right); \Delta_0$\\
\hline
\hline
1/8 & 0.7400 & 0.7656 & 0.6905; 64 & 0.6876; 384 & 0.7355; 16 \\
\hline
1/4 & 0.5894 & 0.5906 & 0.4395; 64 & 0.4454; 448 & 0.5860; 12 \\
\hline
3/8 & 0.4608 & 0.4474 & 0.2440; 64 & 0.2545; 512 & 0.4585; 24 \\
\hline
1/2 & 0.3462 & 0.3281 & 0.1180; 64 & 0.1285; 640 & 0.3445; 28 \\
\hline
5/8 & 0.2427 & 0.2272 & 0.0475; 64 & 0.0556; 832 & 0.2415; 40 \\
\hline
3/4 & 0.1492 & 0.1406 & 0.0135; 64 & 0.0187; 1024 & 0.1480; 52 \\
\hline
7/8 & 0.0665 & 0.0656 & 0.0010; 64 & 0.0030; 448 & 0.0575; 64 \\
\hline
\end{tabular}
\end{table}

\begin{table}[!t]
\renewcommand{\arraystretch}{1.3}
\caption{Results for $q=1024$}
\label{t_1024}
\centering
\begin{tabular}{|c||c|c|c|c|l|}
\hline
R & VG & Upper & $\delta_1\left(R\right); \Delta_0$ & $\delta_2\left(R\right); \Delta_0$ & $\delta_3\left(R\right); \Delta_0$\\
\hline
\hline
1/8 & 0.8036 & 0.7770 & 0.6590; 224 & 0.6319; 192 & 0.8035; 16\\
\hline
1/4 & 0.6573 & 0.5994 & 0.3350; 248 & 0.3217; 276 & 0.6570; 16\\
\hline
3/8 & 0.5252 & 0.4541 & 0.1440; 320 & 0.1374; 304 & 0.5250; 24\\
\hline
1/2 & 0.4028 & 0.3330 & 0.0545; 332 & 0.0524; 384 & 0.4025; 28\\
\hline
5/8 & 0.2884 & 0.2305 & 0.0180; 352 & 0.0170; 384 & 0.2880; 40\\
\hline
3/4 & 0.1817 & 0.1427 & 0.0045; 224 & 0.0045; 640 & 0.1810; 60\\
\hline
7/8 & 0.0835 & 0.0666 & 0.0005; 128 & 0.0005; 768 & 0.0795; 96\\
\hline
\end{tabular}
\end{table}

\section{Conclusion}
{
A new upper bound on the minimum distance of expander codes is derived. The bound lies below the Varshamov-Gilbert bound while $q \ge 32$, hence non-binary expander codes are worse than the best existing non-binary codes. Lower bounds for two ensembles of expander codes are obtained. Both of
\parfillskip=0pt

}
\newpage 
\noindent the bounds lie much below the upper bound. A lower bound for LDPC codes with a Reed--Solomon constituent code is obtained. The bound is very close to the Varshamov-Gilbert bound and lies above the upper bound for expander codes.

\bibliographystyle{IEEEtran}
\bibliography{bibliography}

\end{document}